\documentclass{webofc}
\usepackage[varg]{txfonts}   

\def\be{\begin{equation}}
\def\ee{\end{equation}}

\begin{document}
\title{Diffractive patterns in
deep-inelastic scattering and parton genealogy}

\author{
  \firstname{St\'ephane} \lastname{Munier}\inst{1}\fnsep\thanks{\email{stephane.munier@polytechnique.edu}} 
}

\institute{CPHT, \'Ecole Polytechnique, CNRS, Universit\'e Paris-Saclay, Route de Saclay, 91128 Palaiseau, France
          }

\abstract{%
We report on our recent observation that the occurrence of diffractive patterns in the scattering of electrons off nuclei obeys the same law as the fluctuations of the height of genealogical trees in branching diffusion processes.
}
\maketitle
\section{Introduction}
\label{sec:intro}

\begin{figure}[h]
  \centering
    \sidecaption
  \includegraphics[width=0.6\textwidth,clip]{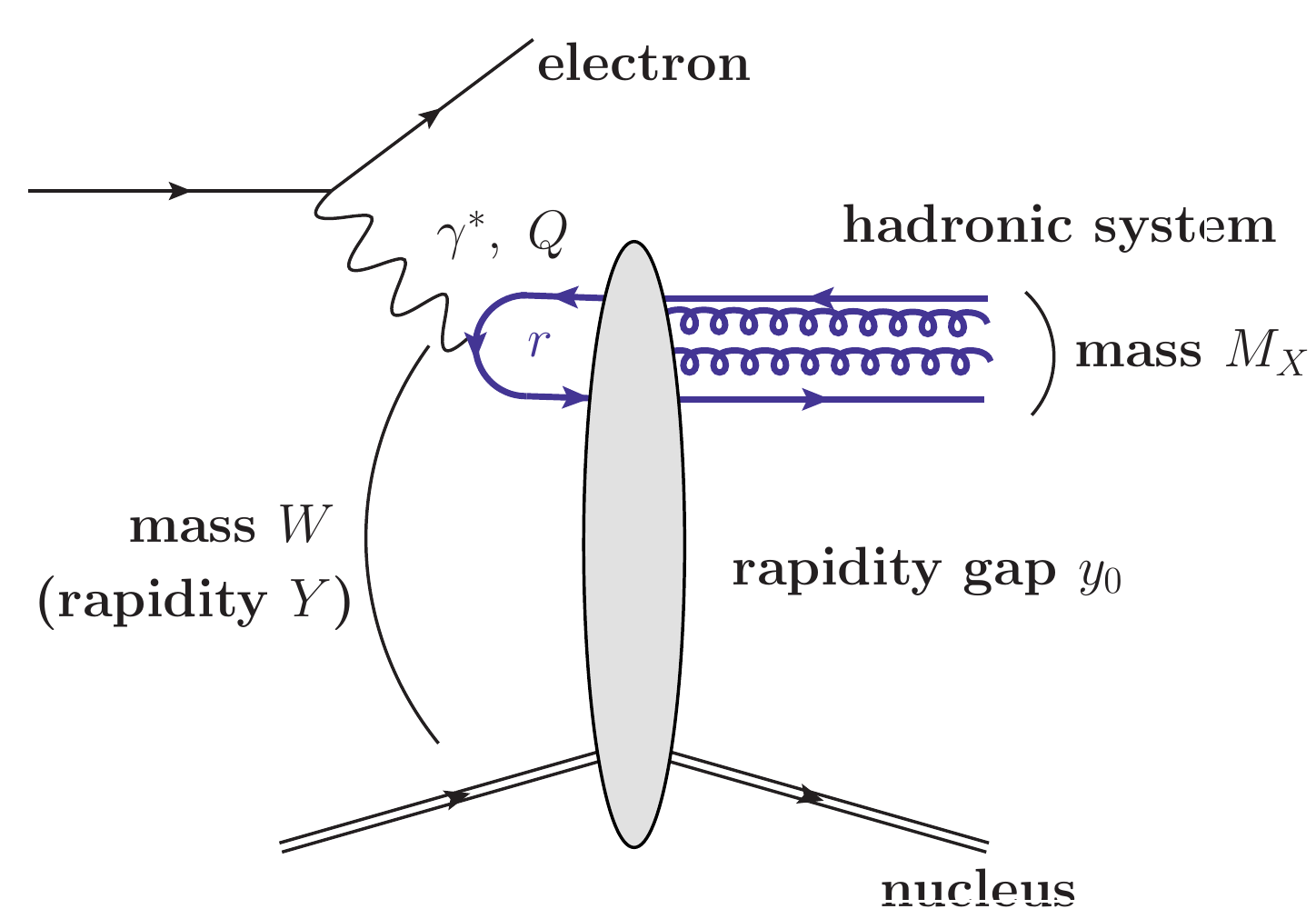}
  \caption{[Adapted from Ref.~\cite{Mueller:2018zwx}]
    Graph contributing to the electron-nucleus diffractive scattering
    amplitude.
    The electron emits a virtual photon,
    which converts to a quark-antiquark pair. The latter scatters elastically, coherently,
    off the nucleus, through a particular quantum fluctuation
    (containing two gluons in addition to the quark and to the antiquark
    in this illustration).
    The final state exhibits a gap of size~$y_0$ here,
    which is a signature of the elastic
    scattering process.
    }
\label{fig-0}
\end{figure}

Consider electrons scattering off atomic nuclei in a high-energy particle collider, a process known as ``deep-inelastic scattering''. Diffractive events are characterized by the striking property that the nucleus is left intact in the process, and is surrounded by an angular region in which no particle is observed, called the ``gap'', the size of which, a quantity fluctuating from event-to-event, is used to classify the diffractive patterns (see Fig.~\ref{fig-0}).

Diffraction in the context of high-energy hadronic and nuclear physics is a surprising phenomenon. Indeed, a nucleus being a lose boundstate of nucleons, themselves being rather fragile compounds of quarks (whenever involved in a reaction of center-of-mass energy larger than a few GeV), one may think that the probability that a nucleus ``survives'' a high-energy collision with a small (compared to nuclear scales) object is negligible.
The observation at the DESY-HERA collider of a significant fraction (about 10\% overall) of diffractive events came as a surprise for many physicists (for a review, see Ref.~\cite{Schoeffel:2009aa}). It boosted the study of quantum chromodynamics (QCD) in the dense, semi-classical, regime, since the most elegant explanation for the important fraction of these events and for their properties came from so-called ``saturation models'', which assumed that at HERA, the proton appeared as a dense system of gluons at momentum scales as high as~1~GeV~\cite{GolecBiernat:1998js}.

Now consider an apparently completely different problem: a particle evolves stochastically in time by diffusing on a line and randomly splitting, at some fixed rate, into two particles, which subsequently follow independently the same rules (see Fig.~\ref{fig-2} below). This process, called ``branching diffusion'', is an interesting mathematical object, potentially relevant in many fields of science, from the physics of glasses to evolutionary biology (for a review by a mathematician, see Ref.~\cite{bovier2016gaussian}). After some given time, pick the two leftmost particles and trace their most recent common ancestor. Surprisingly enough, we found that the distribution of the age of the latter is identical to the distribution of the gap size in diffraction, for a deep reason related to the very mechanism of how fluctuations build up in these two processes.
\begin{figure}[h]
  \centering
    \sidecaption
  \includegraphics[width=0.4\textwidth,clip]{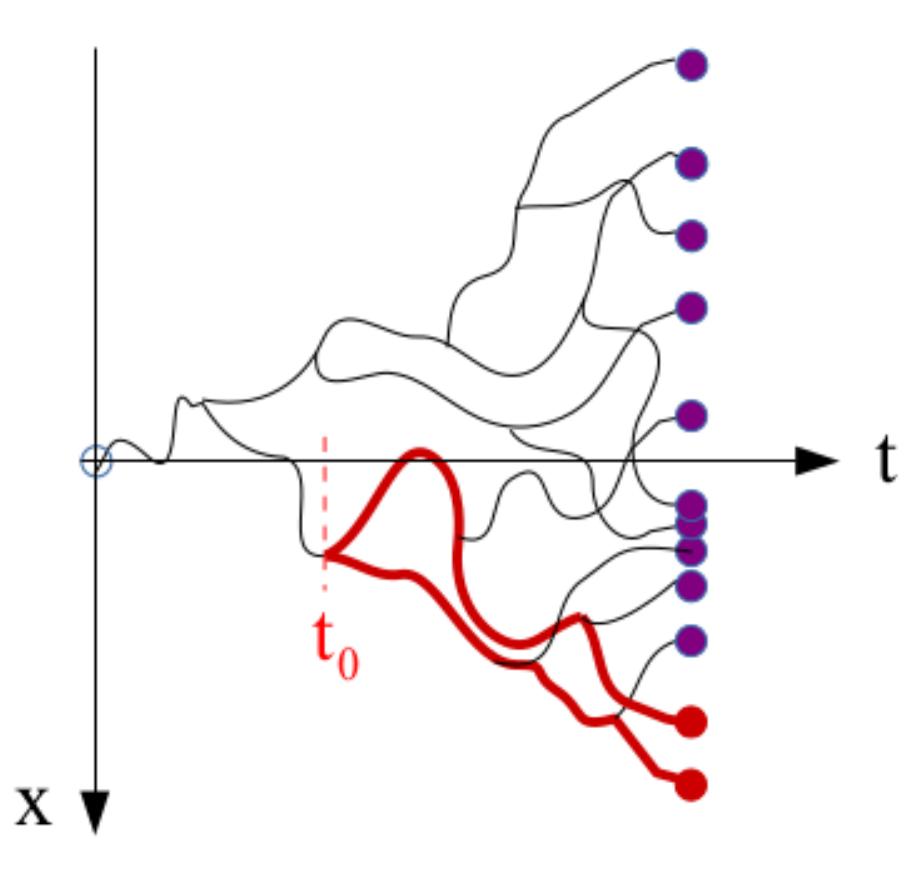}
  \caption{Schematic representation of a realization of a branching Brownian motion process.
    A particle initially at position $x=0$ diffuses, and may split into 2 particles at a constant rate.
    The particles evolve further independently, following the same rules, until the final
    time $t$ is reached. Then, the two particles at the largest positions are picked, and
    their most recent common ancestor is traced: Its decay time is denoted by $t_0$.
    }
\label{fig-2}
\end{figure}

Our study contributes to link distinct fields of science, and paves the way for a better understanding of diffraction phenomena in particle physics, that can be tested at a future Electron-Ion Collider (EIC)~\cite{Accardi:2012qut}.

This short review summarizes the papers in Refs.~\cite{Mueller:2018zwx,Mueller:2018ned}.
We shall start by recalling the equation for the rapidity gap distribution, first derived by Kovchegov and Levin (KL)~\cite{Kovchegov:1999ji}. We then elaborate on a picture which enables us to find asymptotic form of the solution to the latter. We discuss the link with genealogies in branching random walks, before presenting the results of numerical calculations which support our analytical results. The final section gathers some prospects.


\section{\label{sec-2}Equation for the rapidity gap distribution}

Throughout, we shall not address electron-nucleus diffractive scattering, but rather onium-nucleus scattering, which can however be connected to the former through a well-known method. The link is discussed in more detail in Ref.~\cite{Mueller:2018ned}, and is illustrated in Fig.~\ref{fig-0}.

Let us consider the scattering of an onium of size $r$ off a big nucleus at a total center-of-mass
energy corresponding to the rapidity $Y$.
We are going to write down an equation for the evolution of the distribution of the rapidity
gap around the nucleus.

We need to start with the forward elastic onium-nucleus $S$-matrix element. It obeys the Balitsky-Kovchegov (BK)
equation~\cite{Balitsky:1995ub,Kovchegov:1999ua}
\be
\partial_{y} S(r,y)=\bar\alpha\int\frac{d^2r'}{2\pi}\frac{r^2}{r'^2(r-r')^2}\left[
  S(r',y)S(r-r',y)-S(r,y)
  \right],
\label{eq:bkS}
\ee
where the initial condition may be taken from the McLerran-Venugopalan model~\cite{McLerran:1993ni}, which
we can approximate by
$S(r,y=0)=1-e^{-r^2Q_A^2/4}$,
with $Q_A$ the nuclear saturation scale, typically of the order of 1~GeV.
Note that the (dimensionless) total, elastic and inelastic cross sections at a given impact
parameter {\it per unit surface} in the transverse plane to the collision axis may be
expressed with the help of $S$:
\be
\sigma_\text{tot}=2(1-S),\quad\sigma_\text{el}=(1-S)^2,\quad\sigma_\text{in}=\sigma_\text{tot}-\sigma_\text{el}=1-S^2.
\label{eq:sections}
\ee

We now turn to the diffractive cross section
with a given rapidity gap~$y_0$.
According to Kovchegov and Levin,
the distribution of~$y_0$ can be deduced from an auxiliary function $S_2(r,\tilde y)$,
which also obeys the BK equation:
\be
\partial_{\tilde y} S_2(r,\tilde y)=\bar\alpha\int\frac{d^2r'}{2\pi}\frac{r^2}{r'^2(r-r')^2}\left[
  S_2(r',\tilde y)S_2(r-r',\tilde y)-S_2(r,\tilde y)
  \right]\quad\text{for}\quad \tilde y>0,
\label{eq:S2}
\ee
{with the peculiar initial condition}
$S_2(r,\tilde y=0)=\left[S(r,y_0)\right]^2$, which couples the two equations for $S$ and $S_2$.
The gap distribution is then evaluated as
\be
\frac{d\sigma_\text{diff}}{dy_0}=-\frac{\partial}{\partial y_0}S_2(r,Y-y_0).
\ee
These equations also follow quite straightforwardly~\cite{Hatta:2006hs,Mueller:2018ned}
from the Good-Walker picture~\cite{Good:1960ba}.

The solution to the set of equations~(\ref{eq:bkS},\ref{eq:S2})
is actually
what we are after. But the latter are very difficult to solve: there are no known methods to address such
coupled nonlinear evolution equations.
In a certain sense, the work presented here may be viewed as an effort to derive a solution to the KL equations. But instead of trying to solve them brute force,
with purely mathematical tools, we develop a picture of diffractive scattering, from which what we believe should be the large-rapidity asymptotics of the KL equations straightforwardly follow.


\section{Picture of onium-nucleus scattering}

\subsection{Interpretation of the total cross section in different frames}

The total cross section can be deduced from the elastic $S$-matrix element through
the optical theorem, see Eq.~(\ref{eq:sections}). $S$ satisfies the BK equation, the solution of which
is known for some values of the variables, in a limited range:
\be
1-S\simeq \text{const}\times\ln\frac{1}{r^2Q_s^2(Y)}\left[r^2Q_s^2(Y)\right]^{\gamma_0}
\quad\text{with}\quad Q_s^2(Y)\simeq Q_A^2 \frac{e^{\chi'(\gamma_0)\bar\alpha Y}}{(\bar\alpha Y)^{3/2\gamma_0}},
\label{eq:oneminusS}
\ee
where $\bar\alpha\equiv\alpha_s N_c/\pi$, $\gamma_0\simeq 0.63$ is a number, and the values of the
complex function
$\chi(\gamma)=2\psi(1)-\psi(\gamma)-\psi(1-\gamma)$ are the eigenvalues of the kernel of the
linearized BK equation.
This behavior holds when the variables $(r,Y)$ are chosen such that
$1<|\ln [r^2Q_s^2(Y)]|<\sqrt{\chi''(\gamma_0)\bar\alpha Y}$, which we shall call the {\it scaling region}.

$S(r,Y)$ is of course a boost-invariant quantity. But to interpret it, it is useful to choose particular
Lorentz frames.

In the restframe of the {\it onium}, the full QCD evolution is in the nucleus. This evolution
is {\it deterministic} since it starts with a highly-occupied state (the nucleus in its groundstate).
The larger $Y$, the denser the nucleus Fock state: $1-S$, solution to the
BK equation, is an estimator of the density of the nucleus.\footnote{
  Actually, $1-S^2$ is the probability that the onium has an interaction. It may be extracted from the
  data for diffractive elastic vector-meson production~\cite{Munier:2001nr}.}

In the restframe of the {\it nucleus} instead, the QCD evolution is in the onium. Since the latter is a
quark-antiquark pair in its initial (fundamental) state, its evolution is highly stochastic,
and there are large event-by-event fluctuations in its content.
In this context, $1-S$ may be interpreted as the fraction of realizations of the QCD evolution
which lead to a scattering event. We note that there is an interaction between the onium state
and the nucleus whenever there is a gluon in the former which has a transverse momentum
of the order of (or smaller than) the saturation scale of the nucleus, $Q_A$.
Thus $1-S$ can be interpreted as {\it the probability of having a gluon of transverse momentum
  smaller than $Q_A$} in a particular realization of the state of the onium.

\subsection{Total and diffractive cross sections in the
  {\bfseries\bf $y_0$}-frame}

\begin{figure}[h]
  \centering
    \sidecaption
  \begin{tabular}{cc|cc}
\ \ \ \ \ \includegraphics[width=0.27\textwidth,clip]{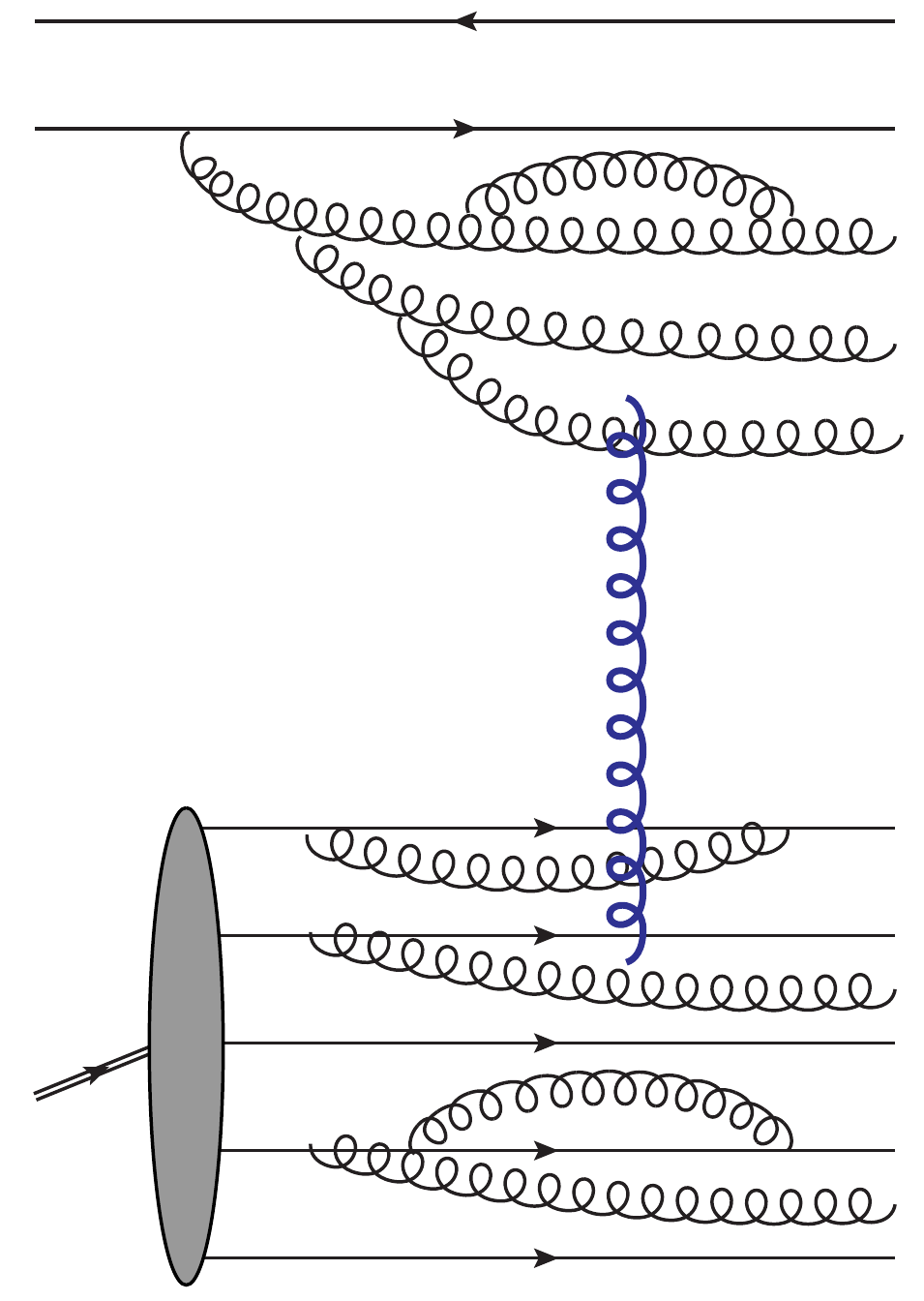}&\ \ \ \ \ \ \ &\ \ \ \ \ \ \ &
  \includegraphics[width=0.396\textwidth,clip]{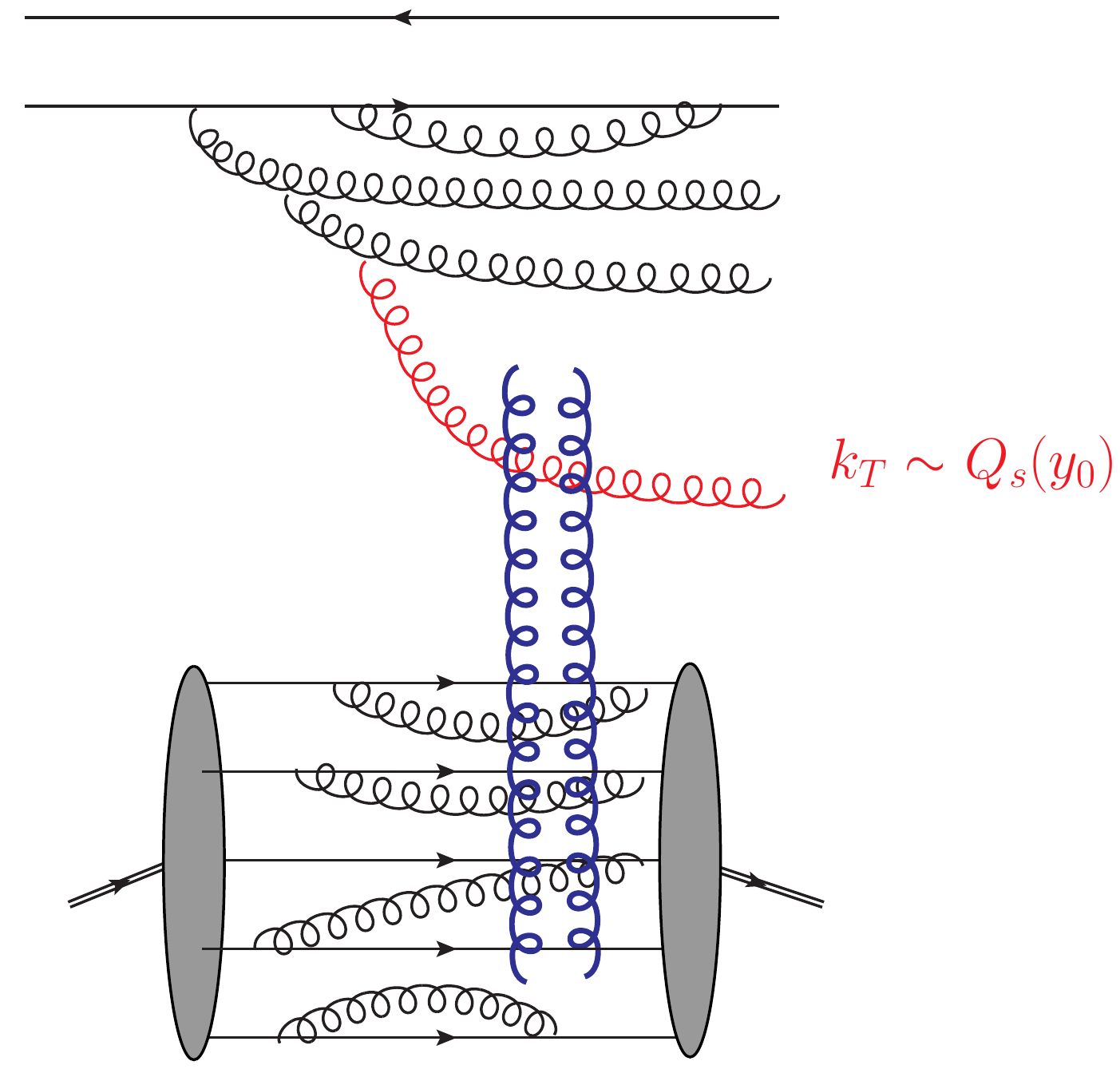}\\
(a)& & & (b)  
  \end{tabular}
  \caption{[Adapted from Ref.~\cite{Mueller:2018zwx}]
    Example of graphs contributing to the total (a) and diffractive dissociation (b)
    amplitude. For the latter, the gap has a size $y_0$, and the condition for its existence
    is the presence of a rare fluctuation, a gluon of transverse momentum $k_\perp\sim Q_s(y_0)$,
  in the Fock state of the onium evolved to rapidity $\tilde y_0=Y-y_0$.}
\label{fig-1}       
\end{figure}

We can also choose an intermediate frame in which the evolution is shared between the
onium and the nucleus. Let us boost the nucleus exactly to the rapidity~$y_0$, while
the onium carries the remaining rapidity $\tilde y_0\equiv Y-y_0$:\footnote{
  The rapidity $y_0$ is also related to the invariant mass $M_X$ of the
  diffracted system: $y_0\simeq \ln W^2/M_X^2$, where $W$ is the total center-of-mass
  energy of the onium-nucleus system, see Fig.~\ref{fig-0}.
}
Examples of
graphs are displayed in Fig.~\ref{fig-1} for the total and diffractive cross section.
In this particular frame, in order to create a rapidity gap, the scattering of the
onium Fock state off the nucleus should be {\it elastic}. For such an elastic scattering
to be probable, there should be
at least a gluon in the state of the onium which has a transverse momentum $k_\perp$ comparable
to (or smaller than)
the characteristic momentum scale of the nucleus in that frame, namely to the
saturation scale $Q_s(y_0)$. We choose the onium size $r$ in the scaling region, in which
the total cross section is small. Therefore,
such a fluctuation is relatively rare: The gluon which interacts is most probably
softer than the typical soft gluons present in these states of the onium. 

So the diffractive cross section conditioned to a gap of size $y_0$,
$d\sigma_\text{diff}/dy_0$,
is proportional to the
probability of having an exceptionally soft gluon, of transverse momentum of the
order of $Q_s(y_0)$, in the Fock state of the onium evolved to rapidity~$\tilde y_0$.
According to the discussion given in the previous section, this quantity
is the solution of the BK equation. Therefore, it reads
\be
\begin{split}
\frac{d\sigma_\text{diff}}{dy_0}&\simeq\text{const}\times
\ln\frac{1}{r^2 \tilde Q_s^2(\tilde y_0)}
\left[r^2 \tilde Q_s^2(\tilde y_0)\right]^{\gamma_0}
\quad \text{with}\quad \tilde Q_s^2(\tilde y_0)=
Q_s^2(y_0)\frac{e^{\bar\alpha\tilde y_0\chi'(\gamma_0)}}{(\bar\alpha\tilde y_0)^{3/2\gamma_0}}\\
&\simeq \text{const}\times\sigma_\text{tot}\left[
  \frac{\bar\alpha Y}{\bar\alpha y_0\,\bar\alpha(Y-y_0)}
  \right]^{3/2}.
\end{split}
\label{eq:rapgap}
\ee
This result is valid in the scaling window\footnote{The region at the border of the scaling window,
  around $|\ln [r^2Q_s^2(Y)]|\sim \sqrt{\chi''(\gamma_0)\bar\alpha Y}$, may be described by Eq.~(\ref{eq:rapgap})
  supplemented with the factor $\exp\left[-\frac{\ln^2 r^2Q_s^2(Y)}{2\chi''(\gamma_0)\bar\alpha (Y-y_0)}\right]$,
  see Ref.~\cite{Mueller:2018ned}.} defined above, and under the additional conditions
$\bar\alpha y_0,\bar\alpha(Y-y_0)\gg 1$.
In order to arrive at the final expression in Eq.~(\ref{eq:rapgap}), we have just used the expression
of the saturation scale as a function of the rapidity given in Eq.~(\ref{eq:oneminusS}).


\section{Ancestry in branching random walks}

It turns out that the structure of diffractive events
bears similarities with an apparently completely different problem: The genealogy of
particles near the boundary of a
branching random walk (BRW) in Ref.~\cite{0295-5075-115-4-40005}.
That dipole evolution (at fixed impact parameter) is a peculiar branching random walk
has been known for some time (see e.g.~\cite{Munier:2014bba}), but that the very structure of the evolution may play
a role is new and quite surprising.

We may define a branching random walk in the following way (see Fig.~\ref{fig-2} for an illustration).
Let us consider a set of particles evolving in time $t$ by diffusing in the real variable~$x$,
and assume that each of them has a probability to split to two particles, independently of
the other particles. (Diffusion may actually be replaced by discrete jumps, which may
occur at the same time as the splitting: the fundamental nature of the problem keeps unchanged).

Assume that the mean density of particles $n(x,t)$ obeys the equation
$\partial_t n=\chi n$; $\chi$ is an appropriate operator acting
on $n$ viewed as a function of $x$. For example, $\chi=\partial_x^2+1$ in the case
of the branching Brownian motion.
$\chi$ admits the eigenfunctions $e^{-\gamma x}$ and we denote by
$\chi(\gamma)$ the corresponding eigenvalues.
After the (large) time~$t$, pick exactly the two leftmost particles,
and look for the first common ancestor splitting time $t_0$ (see Fig.~\ref{fig-2}).

Then, according to Ref.~\cite{0295-5075-115-4-40005},
$t_0$ is distributed as
\begin{equation}
p(t_0)=c_p\left[
  \frac{t}{t_0(t-t_0)}
  \right]^{3/2},
\quad \text{with}\quad c_p=\frac{1}{\gamma_0}\frac{1}{\sqrt{2\pi\chi''(\gamma_0)}},
\label{eq:p}
\end{equation}
where $\gamma_0$ solves $\chi(\gamma_0)=\gamma_0\chi'(\gamma_0)$.
Hence, up to the overall
normalization, which is determined in the case of the genealogies, but
not in the case of diffraction,
$({1}/{\sigma_\text{tot}})({d\sigma_\text{diff}}/{dy_0})$ (see Eq.~(\ref{eq:rapgap}))
corresponds to $p(t_0)$,
with the identifications $\bar\alpha Y\leftrightarrow t$,
$\bar\alpha \tilde y_0\leftrightarrow t_0$.

We think that this formal analogy has actually a deep origin, related to a property
of the common ancestor.
In the same way as in our diffraction calculation,
the common ancestor of the boundary
particles also stems from a fluctuation,
in the form of a particle sent ahead of the \textit{expected}
position of the most extreme particle, occurring
in the course of the evolution, at time $t-t_0$.


\section{Numerical checks}

We stressed in Sec.~\ref{sec-2} that it is difficult to address the Kovchegov-Levin equations analytically.
On the other hand, it is quite straightforward to solve them numerically, since it amounts to
integrating successively two BK equations (in their fixed-impact parameter version).

\begin{figure}[h]
  \centering
  \includegraphics[width=0.7\textwidth,clip]{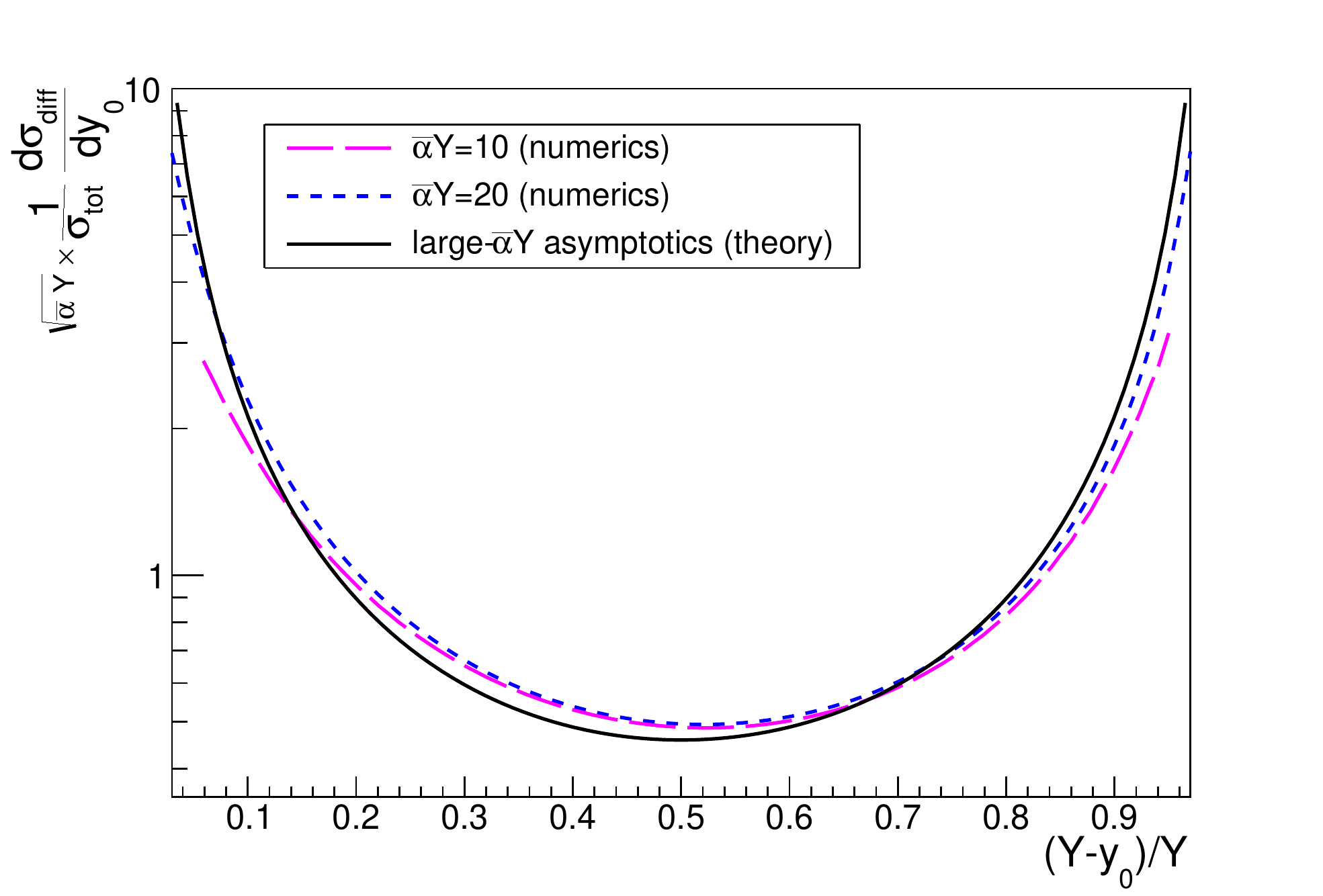}
  \caption{
    [From Ref.~\cite{Mueller:2018zwx}]
    Numerical solution of the KL equation in the asymptotic regime, namely for $\bar\alpha Y=10$
    and $\bar\alpha Y=20$ (see the legend), in order to appreciate the convergence
    to the asymptotics. The results are presented for the diffractive cross section
    normalized to the total cross section, and in scaled variables, in such a way that
    the asymptotic prediction deduced from Eq.~(\ref{eq:rapgap}) (black curve)
    be rapidity-independent. Its normalization is arbitrary.
    }
\label{fig-3}       
\end{figure}

\begin{figure}[h]
  \centering
  \includegraphics[width=0.7\textwidth,clip]{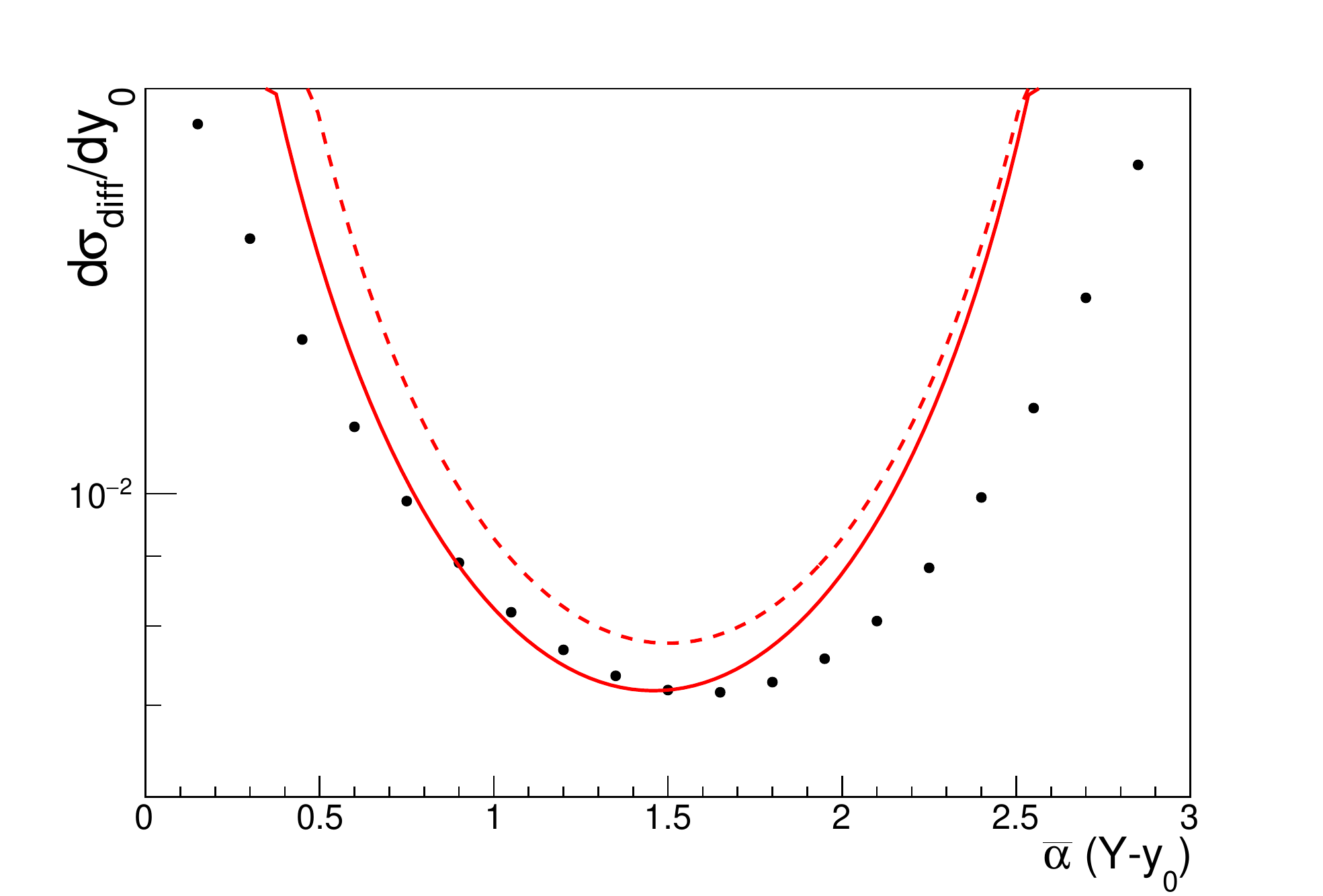}
  \caption{[From Ref.~\cite{Mueller:2018ned}]
    Numerical solution of the KL equation (dots) for $\bar\alpha Y=3$, compared to the asymptotic
    prediction~(\ref{eq:rapgap})
    (dashed line), and to the asymptotics supplemented by a subleading term
    (see Ref.~\cite{Mueller:2018ned} and footnote~3; full line). The
    normalization of the latter is {\it ad hoc}.
    }
\label{fig-4}       
\end{figure}

We first go to very large rapidities in order to see how well the numerical solution of the KL equation
matches our analytical expectations~(\ref{eq:rapgap}).
The comparison is quite spectacular, see Fig.~\ref{fig-3},
and gives us confidence in the validity of our analytical expression.
Note that the KL equation had been solved numerically
before~\cite{Levin:2001yv}, but too low rapidities were considered to
see the asymptotical regime emerging.
For more realistic rapidities, maybe attainable at a future Electron-Ion Collider (EIC),
the rapidity gap distribution still has a similar shape, peaking for
minimal and maximal gap sizes; compare Fig.~\ref{fig-4} to~Fig.~\ref{fig-3}.

\section{Prospects}

On the theory side, we intend to study systematically the asymptotics of the distribution of the rapidity
gap $y_0$, in parallel to the distribution of the decay time of the common ancestor of the largest dipoles
in an event~\cite{Dung:2018}. Preliminary calculations seem to suggest that the overall constant in the gap distribution
may be determined; The first subleading term in the limit $\bar\alpha y_0,\bar\alpha (Y-y_0)\gg 1$ might also be
calculated analytically.

On the phenomenology side, we deem that the distribution of rapidity gaps is an interesting observable to consider
at EIC. Arriving at robust predictions will require to examine more in depth all regimes,
also outside of the
scaling window considered here, and to solve numerically a modified version
of the KL equation which would incorporate
next-to-leading order effects (running coupling~\cite{Kovchegov:2006vj},
kinematical constraint~\cite{Motyka:2009gi},
resummation of collinear logarithms~\cite{Iancu:2015vea}),
known to be quantitatively important.

\begin{acknowledgement}
This work was supported in part by the Agence Nationale
de la Recherche under the project \# ANR-16-CE31-0019.
\end{acknowledgement}

\end{document}